\documentclass[aps,pre,groupedaddress]{revtex4-1}
\usepackage{graphicx}
\begin{document}

\title{Cumulant analysis of the statistical properties of a deterministically thermostated harmonic oscillator}

%\title{Cumulant analysis of the properties of thermostated harmonic oscillator}      
% if too long for running head

%\author{A. N. Artemov}


\author{A. N. Artemov}
\email[]{artemov@fti.dn.ua}
\affiliation{ Donetsk Institute for Physics and Engineering,
Donetsk 83114, Ukraine}


%\maketitle

\begin{abstract}
Usual approach to investigate the statistical properties of
deterministically thermostated systems is to analyze the
regime of the system motion. In this work the cumulant analysis is used to
study the properties of the stationary probability distribution
function of the deterministically thermostated harmonic
oscillators. This approach shifts attention from the investigation
of the geometrical properties of solutions of the systems to the studying a
probabilistic measure. The cumulant apparatus is suitable for
studying the correlations of dynamical variables, which allows one
to reveal the deviation of the actual probabilistic distribution
function from canonical one and to evaluate it. Three different
thermostats, namely the Nos\'e-Hoover, Patra-Bhatacharya and
Hoover-Holian ones, were investigated. It is shown that their
actual distribution functions are non-canonical because of
nonlinear coupling of the oscillators with thermostats. The problem of ergodicity
of the deterministically thermostated systems is discussed.
\end{abstract}

%The sufficient conditions
%for ergodicity were formulated in the terms of cumulants.


\maketitle
\section{Introduction}
\label{intro}

The molecular dynamics approach is widely used to investigate
equilibrium thermodynamic properties of many body systems. The
thermostats, that is physical equations of motion modified by a
dynamic temperature control tool, are commonly used to simulate
systems of particles at constant temperature. There are stochastic
and deterministic thermostats.

In the case of a stochastic thermostat the thermodynamic ensemble is
described by a set of stochastic Langevin equations of motion.
This approach allow us to create a canonical ensemble. But in some
cases this approach is unsuitable because of very slow converging
to the equilibrium state.

Deterministic thermostats are an alternative way to solve the
problem. In this case the ensemble is described by a set of
deterministic nonlinear ordinary differential equations in an
extended phase space. Additional phase variables are one, two or
more pseudo-friction coefficients which obey specific equations of
motion. These equations are arranged in such a way that they can
control some macroscopic parameters of the ensemble. Thus, Nos\'e
\cite{nose} and Nos\'e-Hoover (NH) \cite{hoover} thermostats
enhance the phase space by one variable and control the kinetic
temperature. More complicated  thermostats by Patra-Bhattacharya
(PB) \cite{patra} and Hoover-Holian (HH) \cite{hh} add two
variables and control two parameters of the ensemble. As a result of generalization of these methods in \cite{kus,mart,wat,patra2}, families of thermostats for which the above are special cases were proposed.
Samoletov and Vasiev \cite{samolet} propose a dynamic principle
underlying a range of thetmostats derived using
fundamental laws of statistical physics. Their approach covers
both stochastic and deterministic schemes.

The statistical properties of the stationary dynamical 
system are described by the stationary probability distribution 
function (PDF) the domain of definition of which is the whole phase space.
Such PDF obeys the stationary Liouville continuity equation
corresponding to the system under consideration. For the stationary
dynamical system coupled with any deterministic thermostat one of
the solution of the equation is the PDF which is canonical in the
physical phase space and is Gaussian with respect to the
additional phase variables. But the actual PDF obtained by solving
the equations of motion can be different and it is, of course, not
canonical.

The ergodicity of the system is another problem. The system is
ergodic if its probabilistic measure is invariant or, in other
words, its PDF doesn't depend on initial conditions. In the case
of thermostated systems the usual approach to this problem is to
analyze the regime of the system motion. It boils down to the study
of the  character of the filling of the phase space
with the trajectories of the system.

In the quasi-periodic regime of motion the system evidently is not
ergodic because in this case there exist invariant tori separating
the system phase space into invariant domains depending on the
initial conditions. The chaotic regime is much more consistent with
the concept of ergodicity.

The statistical properties of the deterministically thermostated
systems are investigated mainly on the base of the model of a
one-dimensional harmonic oscillator. The authors of the articles
\cite{hoover,posch,legoll} showed that under some conditions in
the system of the harmonic oscillator coupled with the Nos\'e and NH
thermostats there exist invariant tori. The conditions are
determined by the system parameters and the initial conditions.

A detailed study of the ergodicity of the singly and doubly
thermostated harmonic oscillators in the chaotic regime gave
inconclusive results. The harmonic oscillators coupled with the
Nos\'e and NH thermostats were shown \cite{posch,hoover} to be
nonergodic. Patra and Bhattacharya \cite{patra3}, when working with the
harmonic oscillator coupled with the two-chain NH thermostat, find
that it does not generate the canonical distribution and does not
provide ergodicity of the system. In other papers
\cite{hoover2,hoover3,patra4,hoover4,hoover5} the conclusion was drawn
that this thermostat is ergodic and the ergodicity of some other
thermostats is being discussed.

An undeniable evidence of the nonergodicity of a dynamical system
can be the existence of a conserved integral of motion. Such
integrals are known for the harmonic oscillators thermostated by
NH, HH, NH chain and some other thermostats
\cite{samolet,samolet2}. But, in general, the problem of the
ergodicity of the deterministically thermostated harmonic
oscillator is still open.

In this work, the statistical properties of the harmonic oscillator
coupled with the NH, PB and HH thermostats are studied by means of
the cumulant approach. This method is a powerful tool for studying 
the correlation properties of nonlinear dynamical systems. Its use, 
together with the numerical approach makes it possible to investigate 
the canonicity and ergodicity of the created thermodynamic ensembles.
The minimal information on the cumulant approach, which is
necessary to understand the matter under discussion, is given in
Sec. \ref{sect1}. The statistical properties of the NH system are
considered in Sec. \ref{sect2}. In Sec. \ref{sect3}, the properties
of the PB and HH thermostats are discussed. In the final section,
the conclusive considerations are given. In the Application, one can find some intermediate equations that are mentioned in the text.

\section{Degenerated equations for cumulants.}
\label{sect1}

The purpose of this work is a statistical analysis of deterministic dynamic systems, the evolution of which is determined by the equations of motion. of the form
\begin{equation}\label{e2_1}
    \dot{\textbf{q}}=\textbf{g}(\textbf{q}),
\end{equation}
where $\textbf{q}$ is a vector of phase variables and $\textbf{g}$ is a vector
nonlinear differentiable function. The distribution of the system in
the phase space is described by a PDF $W(\textbf{q},t)$. The
existence of the stationary limit
$W(\textbf{q})=W(\textbf{q},t\rightarrow\infty)$ of this function
is assumed.

The characteristic function $\theta(\textbf{v},t)$ which is a
Fourier transform of the $W(\textbf{q},t)$
\begin{equation}
    \theta(\textbf{v},t)=\int
    e^{\imath(\textbf{qv})}W(\textbf{q},t)d\textbf{q}
\end{equation}
is an equivalent way to describe the distribution of the system.
Moments and cumulants are coefficients of expansion in a series of
the characteristic function and its logarithm respectively
\begin{eqnarray}\label{e2_2}
\nonumber
  \theta(\textbf{v},t)&=& \sum_{m_1,...m_n=0}^\infty
    \frac{\alpha(t)_{m_1,...m_n}^{q_1,...q_n}}{m_1!...m_n!}(\imath
    v_1)^{m_1}...(\imath v_n)^{m_n} \\
  &=&\exp\left[\sum_{m_1,...m_n=0}^\infty
    \frac{\kappa(t)_{m_1,...m_n}^{q_1,...q_n}}{m_1!...m_n!}(\imath
    v_1)^{m_1}...(\imath v_n)^{m_n}\right].
\end{eqnarray}
Here $\alpha(t)_{m_1,...m_n}^{q_1,...q_n}$ is the joint moment and
$\kappa(t)_{m_1,...m_n}^{q_1,...q_n}$ is the the joint cumulant of
$n$ variables $q_1,...q_n$ and $\imath$ is the imaginary unit. The
full set of moments or cumulants completely represents
$W(\textbf{q},t)$ as long as the series (\ref{e2_2}) converge at
all $\{v_i\}_{i=1}^n$.

But the role of the moments and cumulants in the analysis of
statistical, in particular correlation, properties of a dynamical
system is very different. If the joint moment of some dynamical
variables is different from zero, it does not mean the
existence of a statistical dependence of these variables yet. But the
difference from zero of the joint cumulant uniquely manifest such
a dependence.

Hereinafter, the concept of cumulant brackets going to be required.
They are the angle brackets with arguments separated by commas. If
arguments are single variables the cumulant bracket coincides with
the corresponding cumulant, for example $\langle
q_i,q_j,q_i\rangle=\kappa_{2,1}^{q_i,q_j}$. In the case if one or
several arguments are functions the operation "to open cumulant
brackets" is need in order to represent a cumulant bracket as a
function of cumulants \cite{malakh,primak}.

The cumulant analysis is widely used for statistical analysis of
stochastic differential equations in the theory of  Markov
processes \cite{malakh,primak}. To analyze deterministic
differential equations this approach was applied by V. Kontorovich
\cite{kant}. In this work the set of degenerated equations for
cumulants was obtained as the limit of the full set, in which the
amplitudes of random forces tend to zero.

Here we propose simplified derivation of the degenerate equations
for cumulants assuming, that all necessary conditions are
satisfied. The PDF $W(\textbf{q},t)$ corresponding to the system
(\ref{e2_1}) obeys the Liouville continuity equation
\begin{equation}\label{e2_3}
    \frac{\partial W}{\partial t} +
    \frac{\partial \left(\textbf{K}_1 W\right)}{\partial\textbf{q}}
    =
    \frac{\partial W}{\partial t} +
    \frac{\partial \left(\dot{\textbf{q}} W\right)}{\partial\textbf{q}} 
     =0.
\end{equation}
The notation $\textbf{K}_1=\textbf{g}(\textbf{q})$ for kinetic
coefficients is introduced to bring our notations in line with
the established those in the existing literature
\cite{malakh,primak,kant}. Let $f(\textbf{q})$ be a phase
variables function. Then
\begin{eqnarray}\label{e2_4}
 \nonumber
  \frac{d}{dt}\left\langle f(\textbf{q})\right\rangle &=&
  \int f(\textbf{q})\frac{\partial W}{\partial t}d\textbf{q}= \\
  &=&-\int f(\textbf{q})\frac{\partial (\textbf{K}_1W)}{\partial \textbf{q}}d\textbf{q} =
  \left\langle \textbf{K}_1\frac{\partial
  f(\textbf{q})}{\partial\textbf{q}}\right\rangle,
\end{eqnarray}
where the angle brackets mean the statistical average. In the third term, the Liouville equation (\ref{e2_3}) was used to replace the time derivative and the last one is obtained as the result
of the integration by parts.

Substituting monomials of the variables $q_i$ insread of
$f(\textbf{q})$ we arrive at the set of equations for the moments
\begin{eqnarray}\label{e2_5}
 \nonumber
  \frac{d}{dt}\left\langle q_i\right\rangle &=&
  \left\langle K_{1i}\right\rangle, \\
 \nonumber
    \frac{d}{dt}\left\langle q_iq_j\right\rangle &=&
  \left\langle q_i K_{1j}\right\rangle+\left\langle K_{1i}
  q_j\right\rangle =2\left\{\left\langle q_i
  K_{1j}\right\rangle\right\}_s, \\
  \frac{d}{dt}\left\langle q_iq_jq_k\right\rangle &=&
  \left\langle q_iq_j K_{1k}\right\rangle+\left\langle K_{1i}
  q_jq_k\right\rangle+\left\langle q_iK_{1j}q_k
  \right\rangle = \\
   \nonumber
&=&3\left\{\left\langle q_i q_j
  K_{1k}\right\rangle\right\}_s, \\
   \nonumber
&\vdots&
  \end{eqnarray}
  Here the symbol $n\{\ldots\}_s$ is the Stratonovich brackets. It
  denotes the completely symmetrical sum of the variables which are
  enclosed in the brackets. The number before the brackets is the number of terms in
  the expression.

  To obtain the equations for cumulants one should use the
  interrelationship of cumulants and moments. As a result, the
  equations get the same form as for moments if the moment
  brackets (statistical averaging) are replaced by the cumulant
  ones \cite{malakh,primak}.
\begin{eqnarray}\label{e2_6}
 \nonumber
  \frac{d}{dt}\left\langle q_i\right\rangle &=&
  \left\langle K_{1i}\right\rangle, \\
    \frac{d}{dt}\left\langle q_i,q_j\right\rangle &=&
  2\left\{\left\langle q_i,
  K_{1j}\right\rangle\right\}_s, \\
 \nonumber
  \frac{d}{dt}\left\langle q_i,q_j,q_k\right\rangle &=&
  3\left\{\left\langle q_i, q_j,
  K_{1k}\right\rangle\right\}_s, \\
   \nonumber
&\vdots&
  \end{eqnarray}

  In the next sections the stationary statistical properties
  of the different thermostated systems
  are studied. In each case, only a limited number of stationary equations from the infinite set (\ref{e2_6}) are analyzed.

  The phase space dimension of the systems under consideration is three or
  four. The phase variables are the coordinates $x$, the momenta $p$ and
  one or two friction coefficients $\zeta$ and $\xi$,
  $\textbf{q}=(x,p,\zeta(,\xi))$. For the sake of simplicity the
  upper indices in the cumulants are omitted but, instead, the order
  of  the variables in subscripts are fixed, namely

  \begin{equation}\label{e2.7}
    \kappa_{k,l,m(,n)}^{x,p,\zeta(,\xi)}\equiv\kappa_{k,l,m(,n)}.
\end{equation}

\section{Nos\'{e}-Hoover thermostat}
\label{sect2}

\subsection{Equations of motion and canonical PDF}
\label{sect2.1}

Dynamics of a harmonic oscillator which is coupled with the NH
thermostat obeys the set of three ordinary differential equations
of the first order
\begin{eqnarray}\label{e3_1}
 \nonumber
 \dot{x}&=&\frac{p}{m}, \\
 \label{e3.1}
 \dot{p}&=&-kx-\zeta p, \\
 \nonumber
 \dot{\zeta}&=&\frac{1}{\tau}\left(\frac{p^2}{mT}-1\right).
\end{eqnarray}
Here $\textbf{q}=(x,p,\zeta)$ is a vector of phase variables, $m$ and
$\tau$ are the masses of the oscillator and the thermostat, $k$ is
the coefficient of elasticity and $T$ is the temperature.

The well known feature of the system (\ref{e3.1}) is that the
stationary PDF
\begin{equation}\label{e3.2}
    W_c(\textbf{q})\propto \exp\left\{-\frac{p^2}{2mT}-\frac{kx^2}{2T}\right\}
    \exp\left\{-\frac{\zeta^2}{2\tau^{-1}}\right\}
\end{equation}
obeys the stationary Liouville continuity equation
$\frac{\partial}{\partial
\textbf{q}}\left(\dot{\textbf{q}}W(\textbf{q})\right)=0$. The
marginal PDF in the physical phase subspace $(x,p)$ is canonical
and the whole PDF is Gaussian. This function is completely
determined by three nonzero cumulants of the second order
\begin{equation}\label{e3.3}
    \kappa_{2,0,0}=\frac{T}{k},\; \kappa_{0,2,0}=Tm,\;
    \kappa_{0,0,2}=\frac{1}{\tau}.
\end{equation}
All other cumulants are equal to zero. The phase variables,
distribution of which in a phase space are described by such a
function, are statistically independent.

\subsection{Equations for cumulants}
\label{sect2.2}

The kinetic coefficients of the system are
\begin{equation}\label{e3.4}
    \textbf{K}_1=\left(K_{1x}, K_{1p},K_{1\zeta}\right) =
    \left(\frac{p}{m},-kx-\zeta
    p,\frac{1}{\tau}\left(\frac{p^2}{mT}-1\right)\right).
\end{equation}
The equations for cumulants are an infinite set of nonlinear
algebraic equations. In this section the stationary equations
corresponding to nonstationary ones with time derivatives from
$\dot{\kappa}_{1,0,0}$ to $\dot{\kappa}_{0,0,4}$ are
analyzed. The number of the equations is 34. But the key equations
are written out here only. To the left of each expression, 
the corresponding time derivative is written to specify  from
which equation for cumulants this stationary equation was obtained.

The first equation is
\begin{equation}\label{e3.5}
    \dot{\kappa}_{1,0,0}: \;\; \left\langle K_{1x}\right\rangle = \left\langle
    \frac{p}{m}\right\rangle=\frac{1}{m}\kappa_{0,1,0}=0.
\end{equation}
The solution gives a zero value of the cumulant $\kappa_{0,1,0}$.
The second representative equation is more complicated
\begin{equation}\label{e3.6}
    \dot{\kappa}_{0,0,1}: \;\; \left\langle K_{1\zeta}\right\rangle
    = \frac{1}{\tau} \left(\frac{\left\langle p^2\right\rangle}{mT}-1\right)
    = \frac{1}{\tau} \left(\frac{\kappa_{0,2,0}}{mT}-1\right)=0.
\end{equation}
In the last term the dependence of the second moment on cumulants
(\ref{a2}) and the equality to zero of the cumulant
$\kappa_{0,1,0}$ (\ref{e3.5}) are taken into account. Thus, the
equation (\ref{e3.6}) gives $\kappa_{0,2,0}=Tm$. The
characteristic features of these solutions are that they are
completely determined by the equations for cumulants and both of
them coincide with the corresponding cumulants (\ref{e3.3}) of the
canonical PDF (\ref{e3.2}).

There are a number of other zero solutions of the equations
having the same features. They are all cumulants of the first order,
the cumulants $\kappa_{1,1,0}$, $\kappa_{1,0,1}$ and
$\kappa_{0,1,1}$ of the second order, all cumulants of the third
order with the exception of $\kappa_{1,1,1}$, and eight cumulants
of the fourth order.

The rest of the cumulants are the solutions of other type. They
obey the set of equations
\begin{eqnarray}
 \nonumber 
 \dot{\kappa}_{1,1,0}&:&\;\;k\kappa_{2,0,0}+
 \kappa_{1,1,1}-T=0, \\  \nonumber
\dot{\kappa}_{2,0,1}&:&\;\;\frac{1}{\tau T}
\kappa_{2,2,0}+2\kappa_{1,1,1}=0, \\  \nonumber
\dot{\kappa}_{0,2,1}&:&\;\;\frac{1}{\tau Tm} \kappa_{0,4,0} +
\frac{2Tm}{\tau} - 2k\kappa_{1,1,1} - 2Tm\kappa_{0,0,2} =0\\
\label{e3.7}
\dot{\kappa}_{3,1,0}&:&\;\; \frac{3}{m}\kappa_{2,2,0}-k\kappa_{4,0,0}=0, \\
 \nonumber
\dot{\kappa}_{1,3,0}&:&\;\;
\frac{1}{m}\kappa_{0,4,0}-3k\kappa_{2,2,0}- 6Tm\kappa_{1,1,1}=0, \\
\nonumber \dot{\kappa}_{1,1,2}&:&\;\;
k\kappa_{2,0,2}+2\kappa_{1,1,1} \kappa_{0,0,2}-\frac{4}{\tau
}\kappa_{1,1,1}=0
\end{eqnarray}
To obtain these equations the solutions for the canonical
cumulants were taken into account. The set contains six equations
for seven unknowns. So, the set is underdetermined and has an
infinite number of solutions. To solve these equations one need to
take one of the unknowns as a parameter. Let it be
$\kappa_{1,1,1}$. Then
\begin{eqnarray} \label{e3.8.1} %\nonumber
\kappa_{2,0,0} & = & \frac{T}{k}-\frac{1}{k}\kappa_{1,1,1}, \\
\label{e3.8.2} \kappa_{0,0,2}&=&\frac{1}{\tau} +
\frac{1}{T}\left(\frac{3}{\tau}
-\frac{4k}{m}\right)\kappa_{1,1,1}.
\end{eqnarray}
These cumulants differ from canonical ones (\ref{e3.3}) because of
the summands proportional to $\kappa_{1,1,1}$. The fourth order
cumulants, entering into Eqs. (\ref{e3.7}), are directly
proportional to $\kappa_{1,1,1}$. For example
\begin{equation}\label{e3.8.3}
    \kappa_{0,4,0}=6Tm(m-k\tau)\kappa_{1,1,1}.
\end{equation}

Nonzero value of the cumulant $\kappa_{1,1,1}$ means that the
variables $x$, $p$ and $\zeta$ are statistically dependent.
It is easy to understand from the structure of the equations 
for cumulants that the cumulant $\kappa_{1,1,1}$ 
in Eqs. (\ref{e3.7}) arises because of
the nonlinear term $\zeta p$ in the equations of motion.
This means that the coupling of the oscillator with the thermostat,
which is provided by this term, leads to the appearance of
non-physical correlations of the dynamic variables and, thus,
to the non-canonicity of the ensemble.

So, the analysis of the equations for cumulants shows that all
cumulants can be divided into two groups. The first one includes
the cumulants which are unique solutions of the equations. They are
$\kappa_{0,2,0}$ and all zero valued cumulants. These cumulants
coincide with those of the canonical PDF and in what follows are
referred to as canonical. The other group includes the cumulants
which depend on a free parameter, similarly to the solutions of
Eqs. (\ref{e3.7}). These cumulants are nonzero because of the
statistical dependence of phase variables which is the cause of
the distinction between the actual PDF and the canonical one. They
below are termed as non-canonical.

Further analysis of the equations for cumulants gives no any new information. To achieve further progress in studying the
system behavior the equations (\ref{e3.1}) were solved
numerically.

\subsection{Numerical approach}
\label{sect2.3}

The purpose of the numerical approach is to calculate the
cumulants of the stationary PDF as time averages along the 
trajectory of the system motion. But cumulants are not 
statistical averages. They are non-linear combinations of moments,
which are statistical averages. The moments can be calculated 
as time averages asymptotically at $t_{av}\rightarrow\infty$ 
converging to their limiting values
\begin{equation}\label{e3.9}
\langle x^kp^m\zeta^n\rangle_t
    = \frac{1}{t_{av}}\int_t^{t+t_{av}} ds
    \:x(s)^kp(s)^m\zeta(s)^n.
\end{equation}
Here $t_{av}$ is the time of averaging,
% $\alpha^{x\,p\,\zeta}_{kmn}$ is a moment of PDF,
$x(s)$, $p(s)$ and $\zeta(s)$ are the values of the phase
variables on the system trajectory at time $s$. In order 
to calculate the cumulant as a time average, one need
to express it in terms of time averaged moments \cite{malakh,primak} 
and calculate this combination. See for example Eqs. (\ref{a7}, \ref{a8}).

The motion type of the system is determined by its parameters and
initial conditions. The analysis of the system behavior was
performed at fixed parameters $k=m=1$ and $T=0.5$. The thermostat
mass was chosen $\tau=10$ in the regular regime and $\tau\leq2$ in
the chaotic one. The initial conditions are specified later when
the results are discussed. To solve the equations (\ref{e3.1}) the
modification \cite{hockney} of the Verlet algorithm \cite{verlet}
is used with the time step $\triangle t=0.01$ or $0.001$
depending on the parameters.

\subsection{Regular motion}
\label{sect2.4}

\begin{figure} %[h]
  \includegraphics[width=8.5cm]{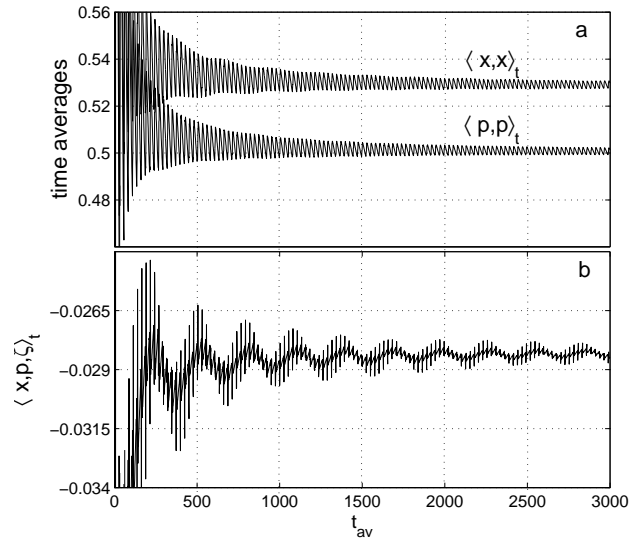}
  \caption{Dependence of the time averages $\langle x,x\rangle_t$,
  $\langle p,p\rangle_t$ (a) and $\langle x,p,\ \zeta\rangle_t$ (b) on
  $t_{av}$ for the NH thermostat in the regular regime of motion. }
  \label{fig.1}
\end{figure}

The regular motion of the system is ensured by rather large value
of the thermostat mass $\tau=10$ and relatively small values of
the initial coordinates $x_0$ and momenta $p_0$. Numerical
calculations show that the expressions to be averaged converge
rather rapidly (the averaging time $t_{av}\sim 10^3-10^4$) to the
corresponding values resulting from the equations for cumulants.
The results of averaging over time of the expressions $\langle
x,x\rangle_t$, $\langle p,p\rangle_t$ and $\langle
x,p,\zeta\rangle_t$ which asymptotically approximate to
$\kappa_{2,0,0}$, $\kappa_{0,2,0}$ and $\kappa_{1,1,1}$ are shown
in Fig.\ref{fig.1} as functions of $t_{av}$ at initial conditions
$x_0=0$, $p_0=0.1$ and $\zeta_0=0$. One can see that these
averages are in good agreement with the expressions (\ref{e3.6})
and (\ref{e3.8.1}).

\begin{figure} %[h]
 \includegraphics[width=8.5cm]{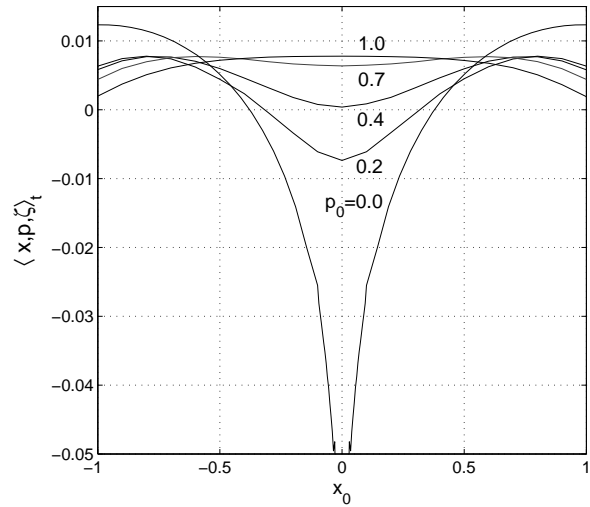}
  \caption{Dependence of the time average $\langle
  x,p,\zeta\rangle_t$
  converging to $\kappa_{1,1,1}$ on the initial coordinate
  $x_0$ at different initial momenta $p_0$ and at the initial
  friction coefficient $\zeta_0=0$. }
  \label{fig.2}
\end{figure}

Special attention was payed to the properties of the cumulant
$\kappa_{1,1,1}$ which in the numerical analysis is approximated
by the time average $\langle x,p,\zeta\rangle_t$. In the previous
subsection this variable was taken as a free parameter and it
cannot be found from the equations. In the numerical approach it
is obtained as the average over time along a trajectory and is
always determined well. Numerical calculation shows that
$\kappa_{1,1,1}$, as well as other cumulants connected with them
(\ref{e3.7}, \ref{e3.8.1}, \ref{e3.8.2}), depends on the initial
conditions. The dependence of $\langle
x,p,\zeta\rangle_t\simeq\kappa_{1,1,1}$ on $x_0$ is shown in
Fig.\ref{fig.2} for $\zeta_0=0$, different $p_0$ and
$t_{av}=3000$.

So, under the condition of the regular motion (motion on torus)
the system PDF is non-canonical and it depends on the initial
conditions, that is the PDF is different on different
trajectories. This means that the system is not ergodic. This
result is in quite agreement with the earlier obtained ones
\cite{hoover,legoll}.

\subsection{Chaotic motion}
\label{sect2.5}

To study the chaotic regime of the system motion the values of the
thermostat mass $\tau\leq 2$ and the initial values of the
oscillator position $x_0=0$ and $0.5$, the momentum $p_0=4.0$, and
$\zeta_0=0$ were chosen. The numerical analysis of the solutions
of Eqs.(\ref{e3.1}) shows that the time averages corresponding to
the canonical and non-canonical cumulants in this case behave
differently.

\begin{figure}%[h]
 \includegraphics[width=8.5cm]{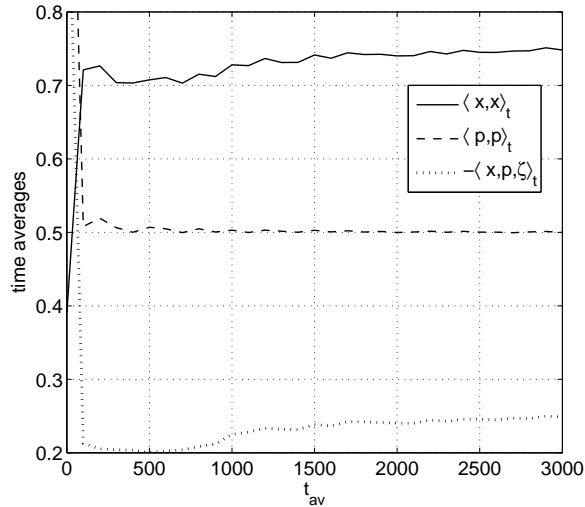}
  \caption{Dependence of the time averages $\langle x,x\rangle_t$,
  $\langle p,p\rangle_t$ and $\langle x,p,\ \zeta\rangle_t$ on $t_{av}$
  for the NH thermostat in the chaotic regime of motion.}
  \label{fig.3}
\end{figure}

The averages approximating to the canonical cumulants converge
rapidly ($t_{av}\sim 10^3-10^4$) to corresponding solutions of the
equations for cumulants. They are $\langle p,p\rangle_t$,
converging to $\kappa_{0,2,0}$, and all averages converging to
zero valued cumulants. As an example see the plot $\langle
p,p\rangle_t$ vs $t_{av}$ in Fig.\ref{fig.3}.

\begin{figure}[h]
 \includegraphics[width=8.5cm]{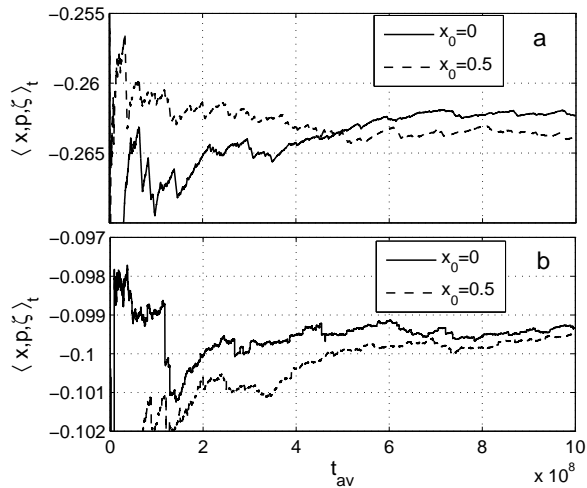}
  \caption{Dependence of the time averages $\langle
x,p,\zeta\rangle_t$,
   on $t_{av}$ at the initial conditions $x_0=0$, $0.5$, $p_0=4$ and $\zeta_0=0$
   for the NH thermostat at $\tau=2$ (a) and $\tau=0.1$ (b) in the chaotic regime of motion.}
  \label{fig.4}
\end{figure}

Other time averages approximating to the non-canonical cumulants
converge to limiting values very slowly. The time average $\langle
x,p,\zeta\rangle_t$ converging to $\kappa_{1,1,1}$ is shown in
Fig.\ref{fig.4} up to $t_{av}=10^9$ for different $\tau$ and
initial conditions. It is clearly seen that the averaged  values
are still far from their limits.

On the other hand all combinations of the time averages
corresponding to the equations for cumulants, including
Eqs.(\ref{e3.7}), converge to their zero values rather quickly
($t_{av}\sim 10^3-10^4$) in spite of the fact that some summands
converge very slowly. An example of such behavior can be found in
Fig.\ref{fig.3} where the averages $\langle x,x\rangle_t$,
$\langle p,p\rangle_t$ and $\langle x,p,\zeta\rangle_t$ are
plotted as functions of $t_{av}$. It is seen that these averages
at $t_{av}=3000$ satisfy the first equation in the set
(\ref{e3.7}) with good accuracy while the individual terms are far
from their stationary values.

In the previous section the dependence of the cumulant
$\kappa_{1,1,1}$ on initial conditions was produced for the case
of periodic motion. This became possible due to the rapid converging
of the time averages. In the case of chaotic motion the
situation is more difficult because of their slow converging. As it can be seen in Fig.\ref{fig.4} that after averaging during time $t_{av}=10^9$ it is still impossible to conclude with certainty whether the limits corresponding to the initial conditions $x_0=0$ and $0.5$ coincide or differ. On the other hand at $\tau=2$
the system demonstrates chaotic motion under the initial
conditions $x_0=0$, $p_0=4$, $\zeta=0$ and periodic one with
$x_0=0$, $p_0=0.4$, $\zeta=0$ that is evidence of the nonergodicity
of the system.

\begin{figure} %[h]
 \includegraphics[width=8.5cm]{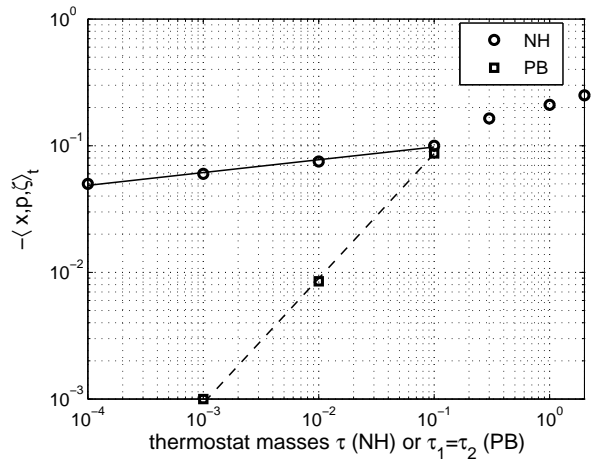}
  \caption{Dependence of the time averages $\langle
x,p,\zeta\rangle_t$
   on the thermostat mass for the NH and PB thermostats in the chaotic regime of motion.}
  \label{fig.4.1}
\end{figure}

In the chaotic regime the dependence of the non-canonical
cumulants (in particular $\kappa_{1,1,1}$) on the thermostat mass
$\tau$ is of special interest. The dependence of $-\langle
x,p,\zeta\rangle_t$ on $\tau$ is shown in Fig.\ref{fig.4.1} with
open circles in the double logarithmic scale. All points are
obtained by the averaging over time and the accuracy of the
obtained values is small. However the points corresponding to
$\tau\leq 0.1$ demonstrate the linear dependence in the
logarithmic scale. This corresponds to the power law which
parameters were found by the mean square method
\begin{equation} %\label{}
    \kappa_{1,1,1}=-0.123\tau^{0.1}.
\end{equation}
It is shown in Fig.\ref{fig.4.1} by a solid line. Such behavior
of the non-canonical cumulants means that the actual PDF, as
$\tau$ decreases, approaches the canonical one.

This feature of the system statistical properties are connected
with the existence of two characteristic times. The first one is
the characteristic time of the mechanical motion of the oscillator
in the phase space which is of the order of $t_{osc}\simeq 2\pi
\sqrt{m/k}$. The second is the characteristic time of the energy
exchange between the oscillator and thermostat and it is
$t_{exch}=\sqrt{\tau}$. So, the conditions for obtaining good
statistical properties of the thermostat is the inequality
$t_{exch}\ll t_{osc}$.

\section{Other thermostats}
\label{sect3}

In this section the statistical properties of the harmonic
oscillator coupled with more complicated thermostats are
discussed. The phase spaces of these systems are four-dimensional.
The extension of the dimension is due to the inclusion of an
additional dynamical variable which is one more friction
coefficient. The advantage of these thermostats is the fact that they ensure
one more cumulant of the actual PDF to be canonical.

The statistical properties of the systems are similar to those
discussed in the previous section. Therefore, they are discussed in
less detail. Only the main traits and distinguishing features of
the systems are represented. Besides, the statistical properties
of the chaotic motion only are discussed.

\subsection{Patra-Bhattacharya thermostat.}
\label{sect3.1}

The dynamical system, which is a harmonic oscillator coupled with
the PB thermostat \cite{patra}, can be described by the set of
four ordinary differential equations
\begin{eqnarray}
 \nonumber
 \dot{x}&=&\frac{p}{m}-\xi x, \\
  \nonumber
 \dot{p}&=&-kx-\zeta p, \\
\label{e4.1}
 \dot{\zeta}&=&\frac{1}{\tau_1}\left(\frac{p^2}{mT}-1\right),\\
 \nonumber
  \dot{\xi}&=&\frac{1}{\tau_2}\left(\frac{kx^2}{T}-1\right).
\end{eqnarray}
Here the phase variables $\textbf{q}=(x, p, \zeta, \xi)$ form a
four-dimencional vector, $\xi$ is the additional configurational
friction coefficient and $\tau_1$ and $\tau_2$ are the masses of
the thermostats.

Gaussian PDF
\begin{equation}\label{e4.2}
    W_c(\textbf{q})\propto \exp\left\{-\frac{p^2}{2mT}-\frac{kx^2}{2T}\right\}
    \exp\left\{-\frac{\zeta^2}{2\tau^{-1}_1}-\frac{\xi^2}{2\tau^{-1}_2}\right\}
\end{equation}
obeys  the stationary Liouville continuity equation corresponding
to this system. The marginal PDF in the physical phase space
$(x,p)$ is canonical and whole PDF is completely determined by
four non-zero cumulants
\begin{equation}\label{e4.3}
    \kappa_{2,0,0,0}=\frac{T}{k},\; \kappa_{0,2,0,0}=Tm,\;
    \kappa_{0,0,2,0}=\frac{1}{\tau_1},\;\kappa_{0,0,0,2}=\frac{1}{\tau_2}.
\end{equation}

The kinetic coefficients of the system are
\begin{eqnarray}
\label{e4.4}
 \textbf{K}_1&=&\left(K_{1x},K_{1p},K_{1\zeta},K_{1\xi}\right)= \\
 \nonumber
 &=&\left(\frac{p}{m}-\xi x,-kx-\zeta p,
 \frac{1}{\tau_1}\left(\frac{p^2}{mT}-1\right), \frac{1}{\tau_2}\left(
 \frac{kx^2}{T}-1\right)\right).
\end{eqnarray}

Only three equations for cumulants, demonstrating the main
features of the actual PDF, are considered in this section. They
are
\begin{eqnarray} \nonumber
     \dot{\kappa}_{0,0,1,0}&:&\;\; \langle K_{1\zeta}
     \rangle=\frac{1}{\tau_1 }\left(\frac{\langle p,p\rangle+\langle p\rangle^2}{m T}-1\right) =0, \\
     \label{e4.5}
     \dot{\kappa}_{0,0,0,1}&:&\;\;\langle K_{1\xi}
     \rangle=\frac{1}{\tau_2}\left(\frac{k\langle x,x\rangle+k\langle
     x\rangle^2}{T}-1\right)=0, \\
     \nonumber\dot{\kappa}_{1,1,0,0}&:&\;\;\langle p, K_{1x}\rangle +
     \langle x,K_{1p}\rangle= \\ \nonumber
     &=&\frac{1}{m}\langle p,p\rangle-\langle p,\xi x\rangle -
     k\langle x,x\rangle-\langle x,\zeta p\rangle=0.
\end{eqnarray}
In the first two equations the averages $\langle x^2\rangle$ and
$\langle p^2\rangle$ are transformed in accordance with Eqs.
(\ref{a1}) and (\ref{a2}). In order to get resulting expressions
one need to open cumulant brackets $\langle x,\zeta p\rangle$
(\ref{a3}) and $\langle p,\xi x\rangle$ (\ref{a4}) and to take
into account that all cumulants of the first order are equal to
zero. The latter fact follows from the symmetry of Eqs.
(\ref{e4.1}) and are verified by the numerical calculations.

As a result, Eqs. (\ref{e4.5}) lead to follow expressions for
cumulants
\begin{eqnarray}
\label{e4.6}
  \kappa_{0,2,0,0} &=& Tm, \\
\label{e4.7}
  \kappa_{2,0,0,0} &=& \frac{T}{k}, \\
\label{e4.8}
  k\kappa_{2,0,0,0}&=&\frac{1}{m}\kappa_{0,2,0,0} - \kappa_{1,1,1,0} -
  \kappa_{1,1,0,1}.
\end{eqnarray}
The expressions (\ref{e4.6}) and (\ref{e4.7}) show that the
cumulants $\kappa_{0,2,0,0}$ and $\kappa_{2,0,0,0}$ are canonical.
This means that the PB thermostat controls both the kinetic and
the configurational temperature in contrast to the NH thermostat
which controls only the kinetic one.

The expression (\ref{e4.8}) together with (\ref{e4.6}) and
(\ref{e4.7}) results in the equality $\kappa_{1,1,1,0}= -
\kappa_{1,1,0,1}$. Numerical analysis shows that these cumulants
are nonzero and the equality is satisfied well at $t_{av}\geq
1000$. This means that the PDF of the PB system is non-canonical
because of the nonphysical statistical dependence of the phase variables.

It is clear from the structure of the equations (\ref{e4.5}) that
non-canonical cumulants $\kappa_{1,1,1,0}$ and $\kappa_{1,1,0,1}$
in Eqs. (\ref{e4.6})-(\ref{e4.8}) arise because of the presence of
nonlinear terms $\zeta p$ and $\xi x$ in the equations of motion (\ref{e4.1})
that ensure the coupling of the oscillator with the thermostat.
So, the origin of the non-canonicity of the PB system is the same as for the NH one,
namely, the nonlinearity introduced into the equations of motion by the thermostat.

\begin{figure}%[h]
 \includegraphics[width=8.5cm]{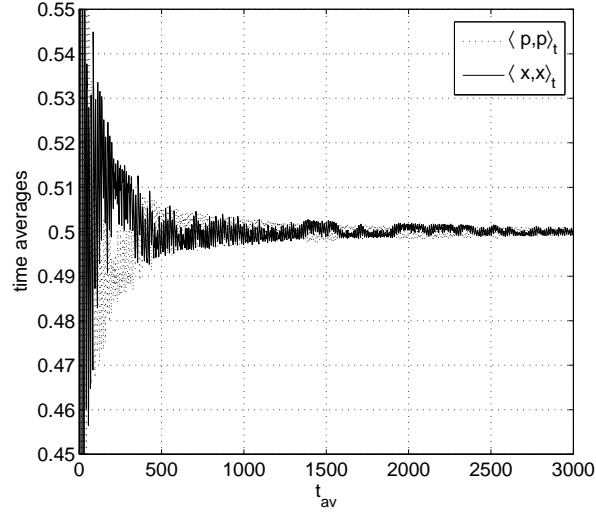}
  \caption{Dependence of the time averages $\langle x,x\rangle_t$,
  and $\langle p,p\rangle_t$ on $t_{av}$ for the PB thermostat at
  the initial conditions $x_0=0$, $p_0=4$, $\zeta_0=0$ and $\xi_0=0$.}
  \label{fig.5}
\end{figure}

The numerical solution of Eqs. (\ref{e4.1}), presented here, was
obtained at the model parameters $k=m=1$, $T=0.5$,
$\tau_1=\tau_2\leq 1$ and the initial conditions $x_0=0$ and
$x_0=0.5$, $p_0=4$, $\zeta_0=0$ and $\xi_0=0$. The rates of
converging of the time averages to the canonical and non-canonical
cumulants, as well as in the case of the NH thermostat,  vary greatly.

\begin{figure}%[h]
\includegraphics[width=8.5cm]{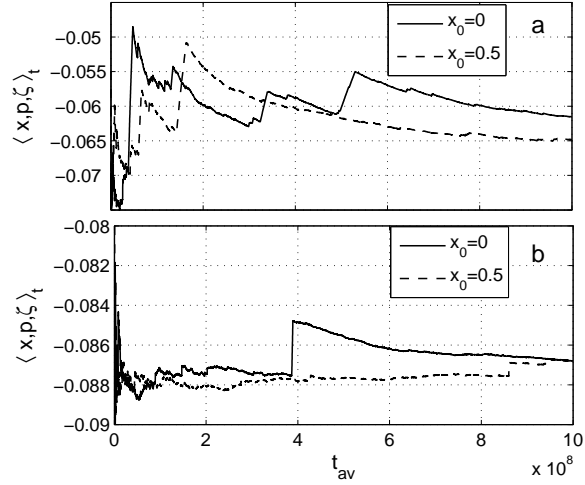}
  \caption{Dependence of the time averages $\langle
  x,p,\zeta\rangle_t$ on
   $t_{av}$ for the PB thermostat at the initial conditions
   $x_0=0$, $0.5$, $p_0=4$, $\zeta_0=0$ and $\xi_0=0$ and at
   $\tau_1=\tau_2=1$ (a) and $0.1$ (b).}
  \label{fig.6}
\end{figure}
The averages $\langle x,x\rangle_t$ and $\langle p,p\rangle_t$
approximating to the canonical cumulants $\kappa_{2,0,0,0}$ and
$\kappa_{0,2,0,0}$ are shown in Fig.\ref{fig.5} as functions of
$t_{av}$. They converge rather rapidly to the configurational and
kinethic temperatures $T=0.5$. On the other hand, the time average
$\langle x,p,\zeta\rangle_t$ plotted in Fig.\ref{fig.6}
 converges to $\kappa_{1,1,1,0}$ much slower.

The dependence of the $\langle
x,p,\zeta\rangle_t\simeq\kappa_{1,1,1,0}$ on the thermostat masses
$\tau_1=\tau_2$ is also established. It is shown in
Fig.\ref{fig.4.1} with open squares. The fitting function
\begin{equation}%\label{}
    \kappa_{1,1,1,0}=-0.837\tau^{0.99}
\end{equation}
is plotted by a dashed line.

\subsection{Hoover-Holian thermostat.}
\label{sect3.2}

The dynamical system which is a harmonic oscillator, coupled with
the HH thermostat \cite{hh}, can be represented by the set of four
ordinary differential equations
\begin{eqnarray}
 \nonumber
 \dot{x}&=&\frac{p}{m}, \\
  \nonumber
 \dot{p}&=&-kx-\zeta p-\xi p^3, \\
\label{e4.9}
 \dot{\zeta}&=&\frac{1}{\tau_1}\left(\frac{p^2}{mT}-1\right),\\
 \nonumber
  \dot{\xi}&=&\frac{1}{\tau_2}\left(\frac{p^4}{m^2T^2}-\frac{3p^2}{mT}\right).
\end{eqnarray}
The canonical Gaussian PDF (\ref{e4.2}) obeys the Liouville
continuous equation corresponding to this system and it is
completely defined by four nonzero cumulants (\ref{e4.3}).

The kinetic coefficients of the system are
\begin{eqnarray}
\label{e4.10}
 \textbf{K}_1&=&\left(K_{1x},K_{1p},K_{1\zeta},K_{1\xi}\right)= \\
 \nonumber
 &=&\left(\frac{p}{m},-kx-\zeta p-\xi p^3, \right.\\
  \nonumber
&&\left.\;\;\;\frac{1}{\tau_1}\left(\frac{p^2}{mT}-1\right),
\frac{1}{\tau_2}\left(
 \frac{p^4}{m^2T^2}-\frac{p^2}{mT}\right)\right).
\end{eqnarray}

The equations for cumulants, which are considered here, are
\begin{eqnarray} \nonumber
     \dot{\kappa}_{0,0,1,0}&:&\;\; \langle K_{1\zeta}
     \rangle=\frac{1}{\tau_1 }\left(\frac{\langle p,p\rangle+\langle p\rangle^2}{m T}-1\right) =0, \\
     \label{e4.11}
     \dot{\kappa}_{0,0,0,1}&:&\;\;\langle K_{1\xi}
     \rangle=\frac{1}{\tau_2}\left(\frac{\langle p^4\rangle}{m^2T^2}- 3\frac{\langle p^2\rangle}{mT}\right)=0, \\
     \nonumber\dot{\kappa}_{1,1,0,0}&:&\;\;\langle p, K_{1x}\rangle +
     \langle x,K_{1p}\rangle= \\ \nonumber
     &=&\frac{1}{m}\langle p,p\rangle-
     k\langle x,x\rangle -\langle x,\zeta p\rangle-\langle x,\xi p^3\rangle=0.
\end{eqnarray}

After opening the cumulant brackets (see Appendix) and taking into
account, that all cumulants of the first order and all joined
cumulants of the second order are equal to zero (verified by
numerical calculations), the solutions of the Eqs.(\ref{e4.11}) take
the form
\begin{eqnarray} \label{e4.12.1}
  \kappa_{0,2,0,0} &=& mT, \\ \label{e4.12.2}
  \kappa_{0,4,0,0} &=& 3\kappa_{0,2,0,0}(mT-\kappa_{0,2,0,0})=0,
  \\ \label{e4.12.3}
  k\kappa_{2,0,0,0} &=& \kappa_{0,2,0,0} \left(\frac{1}{m}-3\kappa_{1,1,0,1} \right)
    -\kappa_{1,1,1,0}- \kappa_{1,3,0,1}.
\end{eqnarray}

Eq. (\ref{e4.12.1}) shows that the HH thermostat, as well as the
NH and PB ones, controls the kinetic temperature of the system.
The expression (\ref{e4.12.2}) demonstrates the fact that the HH
thermostat controls additionally the fourth moment $\langle
p^4\rangle$ (see Eq.(\ref{a5})) in such a way as to ensure its
equivalence to the canonical one. In the terms of cumulants this
condition corresponds to the equality $\kappa_{0,4,0,0}=0$
(compare it with the same cumulant (\ref{e3.8.3}) of the NH system
PDF).

\begin{figure}%[h]
 \includegraphics[width=8.5cm]{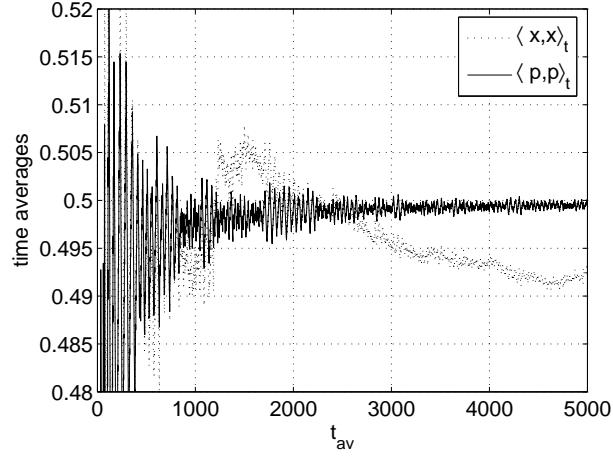}
  \caption{Dependence of the time averages $\langle x,x\rangle_t$
  and $\langle p,p\rangle_t$ on
   $t_{av}$ for the HH thermostat at the initial conditions $x_0=0$, $p_0=4$, $\zeta_0=0$ and $\xi_0=0$.}
  \label{fig.7}
\end{figure}

\begin{figure}%[h]
 \includegraphics[width=8.5cm]{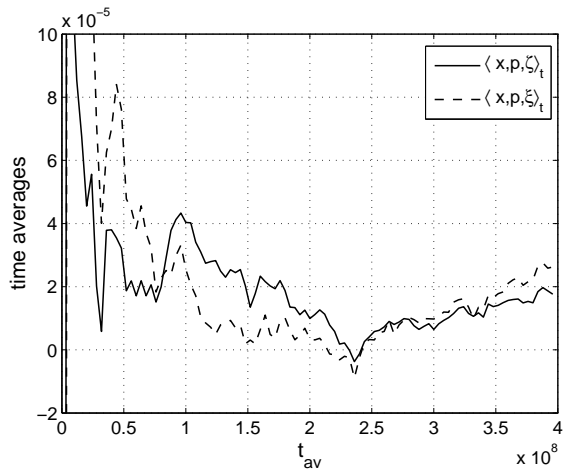}
  \caption{Dependence of the time averages $\langle
  x,p,\zeta\rangle_t$ and $\langle x,p,\xi\rangle_t$ on
   $t_{av}$ for the HH thermostat with masses $\tau_1=\tau_2=0.1$
   at the initial conditions $x_0=0$, $p_0=4$, $\zeta_0=0$ and $\xi_0=0$.}
  \label{fig.8}
\end{figure}

The numerical analysis of the statistical properties of the
dynamical system (\ref{e4.9}) shows that its behaviour is
qualitatively similar to that of the NH and PB thermostats. The
time averages approximating to the canonical cumulants converge
rather rapidly to their equilibrium values (see $\langle
p,p\rangle_t$ in Fig.\ref{fig.7}). But the averages approximating
to the non-canonical cumulants converge much more slowly,
Fig.\ref{fig.8}. The nonzero values of these cumulants, as it was shown in the
previous cases, are the effect of the statistical dependence of the
phase variables because of the nonlinearities $\zeta p$ and $\xi p^4$
introduced into the equations of motion by the thermostat.

\section{Discussion and conclusion}
\label{sect4}

Equilibrium states in statistical mechanics are considered as
canonical ensembles which are described by the Gibbs PDF. This PDF
is determined by the total energy of the system. In most cases
which are of interest the total energy is the sum of the kinetic and
potential energies. Then the Gibbs PDF can be represented as a
product of two functions, one of which depends only on momenta,
and the other only on spatial coordinates. This means that the
coordinates and momenta in the canonical ensemble are
statistically independent.

The deterministic thermostats introduce additional nonlinearity
into the equation of motion of the system. These nonlinearities
lead to the appearance in the ensemble of nonphysical statistical
dependencies between dynamical variables. In this work the analysis of
these dependencies is represented on the example of three thermostats.
The cumulant approach  used is well suited for solving just this kind of
problems.

The properties of the PDF of the harmonic oscillator, coupled with
the NH, PB and HH thermostats, were analyzed. It is shown that
the PDFs of the systems considered are non-canonical. This is
revealed in  the fact that some cumulants, which are equal to zero
in the canonical PDF, are different from zero. This is a
consequence of the statistical dependence of the phase variables
because of the nonlinearities $\zeta p$, $\xi x$ and $\xi p^3$ introduced into
the equations of motion by the thermostats. The degree of
the non-canonicity of the PDFs studied can be evaluated by the
value of the non-canonical cumulants, such as $\kappa_{1,1,1}$ for
the NH thermostat or $\kappa_{1,1,1,0}$ and $\kappa_{1,1,0,1}$ for
the PB and HH ones.  For the NH and PB thermostats it is shown that the thermostat masses reduction leads to decreasing the values of non-canonical cumulants and, therefore, to approximating the actual PDF to the canonical one.

The ergodicity is more difficult to analyze. In the case of the
quasiperiodic motion the situation is clear. The invariant tori
exist in this case and the PDF of the system depends on the initial
conditions because of such dependence of the non-canonical
cumulants. This means that the system is nonergodic.

The chaotic regime of systems is more in line with the idea of ergodicity. But the cumulant method doesn't allow
to identify directly the ergodicity or nonergodicity of the
system. It gives only sufficient conditions of ergodicity. If the
set of equations for cumulants is well determinated, i.e. all
equations are independent and the set is compatible, the system is
ergodic. In this case the PDF is completely determined by the
system parameters and doesn't depend on the initial conditions for
the equation of motion.

It should be noted that now it cannot be ruled out the possibility
that these conditions are also necessary. But this statement has
to be proved.

The systems that are analyzed in the work, do not meet these conditions.
As it was shown in Sec.\ref{sect2}, the set of equations for
cumulants of the NH system splits into two subsets. One of the subsets is
quite determinated and the other is underdetermined. The solutions
of the first subset are the canonical cumulants and those of the
second one are non-canonical. So, the whole set is underdetermined
and the PDF of the system is non-canonical and, most likely,
non-ergodic.

The PB and HH systems were studied in a lesser extent. But existence
of the rapidly and slowly converging time averages to the
canonical and non-canonical cumulants, analogous to that in the NH
system, is indirect evidence in favour of the fact that the
structure of the sets of the equations for cumulants are also
analogous.

The existence of different rates of convergence of time averages,
corresponding to canonical and noncanonical cumulants, also allows one to draw some conclusions
about the ergodicity of the systems under discussion.
To do this one have to take into account the difference in the behaviour of the regular and
chaotic trajectories of the systems and the approximate nature of the numerical solution of
the differential equations of motion of the systems.

A characteristic feature of periodic and quasiperiodic motions is that trajectories close at the initial time instant $t=0$ (close initial conditions) remain close for an arbitrarily long time.  Chaotic trajectories behave differently. Trajectories that are arbitrary close at some point in time can diverge to a large distance (of the order of
the characteristic system dimension) during a finite time approaching at the same time the trajectories with different initial conditions. This instability has been intensively studied using the Lyapunov exponent spectrum, see \cite{licht,eckmann,thompson} and references therein.

A special feature of the numerical approach to the analysis of dynamical systems is that the differential
equations of motion are replaced by difference equations with a finite time step $dt$.
The solutions of such equations coincide with the differential ones only
approximately. Their accuracy depends on the difference scheme used and the time step value.
This means that the numerical solution at each time step jumps from one trajectory at time
$t$ to another close one at time $t+dt$.

In the case of regular motion of the system the numerical trajectory occupies a small tube,
thickness of which is determined by the method accuracy, around the exact trajectory. Since all exact
trajectories that are placed in the tube correspond to close initial conditions, the averages
calculated along the numerical one can be find with the required accuracy irrespective of their
dependence on the initial conditions. An example of such calculations is shown in subsection \ref{sect2.4}.

In the case of chaotic motion the situation is very different. A numerical trajectory in this
case is not close to any exact one. It represents a set of points placed on different
trajectories with very different boundary conditions. The averaging along such a trajectory is
not the same as averaging along any exact one. If the value under calculation does not depend on
the initial conditions, then it does not matter over which set of points the corresponding
expression is averaged, the result will be achieved during rather short time and will be well
determined. Canonical cumulants satisfy these conditions. Examples of the time averages converging to canonical cumulants can be seen in Figs.\ref{fig.3}, \ref{fig.5}, \ref{fig.7}.
Otherwise, when the desired value depends on the initial conditions, the result of
averaging along the numerical trajectory is not related to the time average along any exact
trajectory. Under this condition the time averages of the corresponding expressions "converge"
much slower. This kind of behavior typical of non-canonical time averages can be seen in Figs.\ref{fig.4}, \ref{fig.6}, \ref{fig.8}. The behavior of the time averages described here
is an indication that the PDFs of the NH, PB and HH thermostats depend on the initial conditions,
and consequently the thermodynamic ensembles created by these thermostats are non-ergodic.
Therefore, one can conclude that the canonicity and ergodicity of these systems are closely connected with each other, that is, the non-canonicity of the ensemble means its non-ergodicity.

Generalizing all the above-stated, one can conclude that not only the NH, PB and HH thermostats,
but all deterministic thermostats create non-canonical and non-ergodic ensembles.
The results obtained can also be qualitatively extended to nonlinear and many body systems.
As it is shown above, non-physical correlations of dynamic variables, which are an attribute
of an ensemble non-canonicity and, apparently, non-ergodicity, are determined by the presence
of nonlinear terms in the equations of motion that provide a coupling of the system
with the thermostat. The type of thermostat and its parameters affect
only the degree of non-canonicity.

So, the cumulant approach is proposed as a new method for the studying the statistical properties
of deterministically thermostated dynamical systems. In particular, it allows one to test
the canonicity and ergodicity of the thermodynamical ensembles created by thermostats and to evaluate
the degree of their non-canonicity. The applicability of the method is not limited only to
the harmonic oscillator. It can be successfully used to investigate nonlinear and many-body systems.
It is possible because there is no need to write out and analyze a large number of equations for
cumulants to clarify the main statistical properties of the system. As it was shown by the example
of the PB and HH thermostats, it is enough to study (numerically) only a small number of key cumulants.

\appendix*
\section{The expressions that are referenced in the text.}
\label{app}

The opening of the cumulant brackets used in the text.
\begin{eqnarray}
\label{a1}
  \langle x^2\rangle &=& \kappa_{2,0,0(,0)}+\kappa_{1,0,0(,0)}^2,
  \\ \label{a2}
  \langle p^2\rangle &=& \kappa_{0,2,0(,0)}+\kappa_{0,1,0(,0)}^2,
  \\ \label{a3}
  \langle x,\zeta p\rangle &=& \kappa_{1,1,1,0}+\kappa_{1,1,0,0}\kappa_{0,0,1,0} +
  \kappa_{1,0,1,0}\kappa_{0,1,0,0},
  \\ \label{a4}
  \langle p,\xi x\rangle &=& \kappa_{1,1,0,1}+\kappa_{0,1,0,1}\kappa_{1,0,0,0} +
  \kappa_{1,1,0,0}\kappa_{0,0,0,1},
  \\
  \langle p^4\rangle &=& \kappa_{0,4,0,0}+3 \kappa_{0,2,0,0}^2 + 4\kappa_{0,1,0,0}\kappa_{0,3,0,0}
  \nonumber
  \\ \label{a5}
  &+& 6\kappa_{0,1,0,0}^2\kappa_{0,2,0,0} + \kappa_{0,1,0,0}^4,
   \\
  \langle x,\xi p^3\rangle &=& \kappa_{1,3,0,1} +
  3\kappa_{1,2,0,1}\kappa_{0,1,0,0} +
  \kappa_{1,3,0,0}\kappa_{0,0,0,1} \nonumber \\
  &+& 3\kappa_{1,1,0,1}\kappa_{0,1,0,0}^2 +
  3\kappa_{1,2,0,0}\kappa_{0,1,0,0}\kappa_{0,0,0,1} \nonumber \\
  &+& \kappa_{1,0,0,1}\kappa_{0,1,0,0}^3 +
  3\kappa_{1,1,0,0}\kappa_{0,1,0,0}^2\kappa_{0,0,0,1} \nonumber \\
  \label{a6}
  &+&3\kappa_{1,0,0,1}\kappa_{0,2,0,0}\kappa_{0,1,0,0} +
  3\kappa_{1,1,0,0}\kappa_{0,1,0,1}\kappa_{0,1,0,0},
\end{eqnarray}

The example of the cumulants expressed in terms of the moments
\begin{eqnarray}
  \label{a7}
  \kappa_{0,2,0}&=&
  \langle p,p\rangle = \langle p^2\rangle-\langle p\rangle^2,  \\
  \kappa_{1,1,1}&=&
  \langle x,p,\zeta\rangle = \langle xp\zeta\rangle-\langle
  x\rangle\langle p\zeta\rangle - \langle p\rangle\langle
  x\zeta \rangle \nonumber \\
  &-&\langle \zeta\rangle\langle xp\rangle + 2\langle
  x\rangle\langle p\rangle\langle \zeta\rangle. \label{a8}
\end{eqnarray}

\begin{acknowledgements}
The author thanks A. Samoletov for interesting and useful
discussions.
\end{acknowledgements}

\end{document}